# Bias dependence of tunneling magnetoresistance in magnetic tunnel junctions with asymmetric barriers

Alan Kalitsov,[1#] Pierre-Jean Zermatten,[2] Frédéric Bonell,[3] Gilles Gaudin,[2] Stéphane Andrieu,[3] Coriolan Tiusan,[3,4] Mairbek Chshiev,[2] and Julian P. Velev[1,2,5†]

[1] *Department of Physics, Institute for Functional Nanomaterials, University of Puerto Rico, San Juan, Puerto Rico 00931, USA*
[2] *SPINTEC, UMR (8191) CEA/CNRS/UJF/Grenoble INP, INAC, 17 rue des Martyrs, 38054 Grenoble Cedex, France*
[3] *Institut Jean Lamour, UMR 7198, CNRS-Nancy Universite, BP 239, 5406 Vandoeuvre, France*
[4] *Centre for Superconductivity, Spintronics and Surface Science (C4S), Technical University of Cluj-Napoca, Str. Memorandumului nr. 28, 400114 Cluj-Napoca, Romania*
[5] *Department of Physics and Astronomy, Nebraska Center for Materials and Nanoscience, University of Nebraska, Lincoln, Nebraska 68588,USA*

Email: [#]kalitsov@yahoo.com, [†]jvelev@gmail.com

The transport properties of magnetic tunnel junctions (MTJs) are very sensitive to interface modifications. In this work we investigate both experimentally and theoretically the effect of asymmetric barrier modifications on the bias dependence of tunneling magnetoresistance (TMR) in single crystal Fe/MgO-based MTJs with (i) one crystalline and one rough interface and (ii) with a monolayer of O deposited at the crystalline interface. In both cases we observe an asymmetric bias dependence of TMR and a reversal of its sign at large bias. We propose a general model to explain the bias dependence in these and similar systems reported earlier. The model predicts the existence of two distinct TMR regimes: (i) tunneling regime when the interface is modified with layers of a different insulator and (ii) resonant regime when thin metallic layers are inserted at the interface. We demonstrate that in the tunneling regime negative TMR is due to the high voltage which overcomes the exchange splitting in the electrodes, while the asymmetric bias dependence of TMR is due to the interface transmission probabilities. In the resonant regime inversion of TMR could happen at zero voltage depending on the alignment of the resonance levels with the Fermi surfaces of the electrodes. Moreover, the model predicts a regime in which TMR has different sign at positive and negative bias suggesting possibilities of combining memory with logic functions.

## I. Introduction

The field of spintronics has been very successful in producing magnetoresistive devices for magnetic memory and sensor applications.[1] Magnetic tunnel junctions (MTJs) came to the forefront of spintronics research after theoretical predictions of very high positive tunneling magnetoresistance (TMR) in Fe/MgO/Fe MTJs.[2] Shortly after TMR in excess of 200% was reported experimentally in these junctions.[3,4] More recently TMR as high as 604% at room temperature was reported in MgO-based MTJs with CoFeB electrodes.[5] It has been recognized that the interfaces are crucial for the TMR amplitude and voltage dependence and consequently interface engineering has received a great deal of attention.[6,7]

There are a number of experimental studies of the influence of modified interfaces on the sign and bias dependence of TMR, in particular with insulating layers such as $Ta_2O_5$[8] and NiO,[9] metallic layers Cr[10] and $Fe_3O_4$,[11] adatoms C[12,13] and O,[14,15] and morphologically different interfaces.[16-19] A recent experimental work shows that the insertion of a thin NiO layer at one of the interfaces of a CoFe/MgO/CoFe(001) MTJs gives rise to an asymmetric bias dependence of the TMR.[9] Also switching from positive to negative TMR was observed at larger bias. It was suggested that the effect is due to formation of a non-collinear magnetic structure at the CoFe/NiO interface, however, it is not clear how that fact may affect the bias dependence of TMR. Very similar observations were reported earlier in $NiFe/Ta_2O_5/Al_2O_3/NiFe$ MTJs.[8] The sign reversal was interpreted, using the Jullière model, in terms of the change of the spin polarization of the electrodes with the bias.[20] Strongly asymmetric TMR bias dependence was also reported. Moreover, essentially identical behavior and sign reversal of TMR was observed in experiments doping the interface with non-magnetic metal layers (Cr),[10] non-magnetic atoms (C),[12,13] or by just varying the morphology of the interface.[16]

First principles transport calculations with finite bias are fairly difficult and therefore not commonplace. There are several density functional based calculations of Fe/MgO-based MTJs with ideal and oxidized interfaces at finite bias.[18,21-24] Some of them consider MTJs with modified interfaces and report asymmetric behavior of TMR.[18,20,23] Overall, the asymmetric bias dependence and sign change of TMR emerge as general features of many diverse systems, however, the its interpretation is usually limited to qualitative arguments based on the Jullière model on a case-per-case basis. Further analysis is needed to understand the underlying physics of these phenomena.

In this paper, we report measurements of the bias dependence of TMR in pairs of single crystal Fe/MgO MTJs – one with clean interfaces (Fe//MgO/Fe) and another with a monolayer of O deposited at one of the interfaces (Fe/O/MgO/Fe). Due to the growth procedure, in both cases the bottom (first) interface is atomically sharp while the top (second) is of lower quality.[25] The experiments show that TMR is asymmetric with respect to the voltage sign and negative TMR arises at high voltage. The insertion of an O layer at the higher quality interface acts as an additional barrier and slightly reduces the asymmetry of the bias dependence of TMR.[14,25] We propose a model which explains these experimental observations as well as other previously reported experiments. The model relates the interface asymmetry to the interface transmission functions. We show that interface modifications which preserve the tunneling regime lead to asymmetric bias dependence and TMR inversion at large bias. While modifications with metallic layers can lead to TMR inversion even at zero bias due to resonant transmission.

**II. Experimental results**

The details of the experimental procedure have already been discussed in previous papers.[14,17] The samples were grown on MgO(001) substrate and the MgO barrier thickness was between 1.1 and 2.3 nm (5-11 monolayers). Pairs of samples, Fe//MgO/Fe (*A*) and Fe/O/MgO/Fe (*B*), were grown simultaneously by molecular beam epitaxy in the same ultrahighvacuum (UHV) chamber (base pressure less than $10^{-10}$ Torr). After the deposition of the bottom Fe(001) electrode on both samples, sample *A* was put away in a secondary UHV chamber, adjacent to the growth chamber. Then, molecular O was adsorbed at room temperature on the bottom Fe surface of sample *B* only. The deposition was stoped after adsorption of one O monolayer. The O adsorbtion was controlled in real time *in-situ* by X-ray photoelectron spectroscopy (XPS) as described in detail in the previous work.[14] Then, both samples were annealed at 925K. In sample *B*, the annealing resulted in p(1×1) ordering of the adsorbed O monolayer. Subsequently the MgO barrier and the top Fe layer were grown but not annealed, which resulted in the second MgO/Fe interface to be of lower quality. The annealing of the top electrode was not performed in order to prevent any further evolution of the bottom interface with respect to the initial configuration. The most important feature of this procedure is that the thicknesses, growth, and annealing conditions for both bottom Fe electrodes and the MgO barrier were strictly identical for both A and B samples. Therefore, the differences in the transport characteristics are only related to the presence of the additional O monolayer at the bottom Fe/MgO interface.

Measurements of the bias dependence of TMR were performed on both Fe//MgO/Fe and Fe/O/MgO/Fe MTJs for 1.1 nm, 1.6 nm and 2.3 nm of MgO barrier thickness. The bias dependence of the TMR in the sample with 1.1 nm of MgO barrier is shown in Fig. 1. The maximum TMR value is obtained around zero bias and TMR monotonously decreases with the bias. TMR increases with the thickness of the MgO barrier, with a maximum value of 50% for 1.1nm, 117% for 1.6nm and 142% for 2.3 nm for the MTJs with non-oxidized interfaces. The maximum TMR value obtained for the thicker MgO barrier is slightly lower than the ratio measured earlier,[25] because in these samples the top Fe electrode has not been annealed and therefore the structural quality of the MTJ stack is slightly reduced. Comparing samples A and B for the same MgO barrier, we observe that the presence of one monolayer of O at the bottom Fe/MgO interface decreases the TMR in all cases with respect to the O-free samples (to 38% for 1.1nm, 50% for 1.6nm, and 115% for 2.3nm barriers, respectively). These observations are consistent with previous reports.[2,26,27]

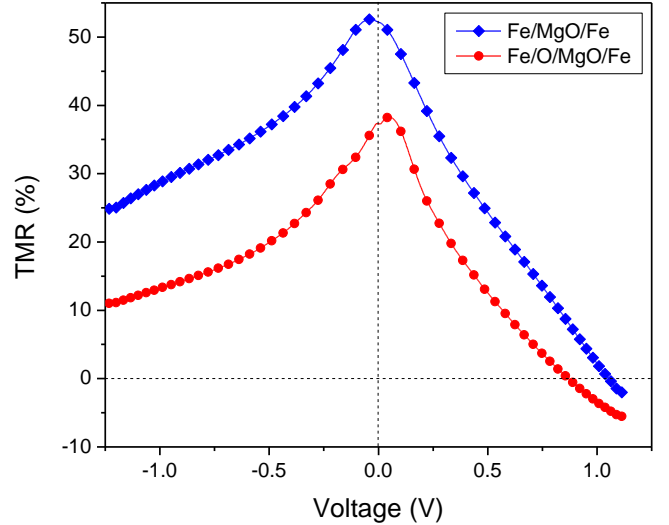

**Figure 1**: Experimental bias dependence of TMR in single crystal Fe//MgO/Fe and Fe/O/MgO/Fe MTJs. The thickness of MgO barrier is 1.1 nm. The top Fe electrode is taken as reference while the bias is applied to the bottom Fe electrode. A numerical smoothing procedure was applied to reduce the experimental noise around zero bias.

In all cases TMR is asymmetric with respect to the voltage. We also observe that TMR changes sign at large bias from positive to negative. Due to the asymmetry, the inversion happens only at positive bias for the experimental range of voltages, which is limited to prevent electric breakdown. In order to quantify and compare the TMR asymmetry between samples, we define $V_+$ and $V_-$ to be the absolute values of the bias at which the TMR reaches the half of its maximum value for positive and negative bias, respectively. For the 1.1nm thick MgO barrier (Fig. 1) with non-oxidized interfaces we obtain $V_+ = 0.42\text{V}$ and $V_- = 1.04\text{V}$. We define the asymmetry coefficient as follows $\gamma = |(V_+ - V_-)/V_-| = 60\%$. This asymmetry is large suggesting that the quality of the two interfaces is markedly different. This is not surprising having in mind that the bottom Fe/MgO interface is atomically sharp after the annealing of the bottom Fe electrode, while top MgO/Fe interface is rougher for couple of reasons. First, the 1.1nm MgO thickness is larger than the critical thickness for the pseudomorphic MgO growth on Fe above which the plastic relaxation occurs,[28] creating dislocations within the MgO. Second, the top Fe electrode is not annealed which reduces its morphological and crystallographic properties compared with the bottom one. The same measure for the oxidized sample gives $\gamma = 28\%$ ($V_+ = 0.29$V and $V_- = 0.4$V). Clearly, the O monolayer reduces the asymmetry of the bias dependence of the TMR.

Similarly, for the MTJs with 1.6nm of MgO for the clean interface we obtain asymmetry of $\gamma = 69\%$ ($V_+ = 0.42$V and $V_- = 1.37$V), while again the oxidized interface makes TMR more symmetric $\gamma = 25\%$ ($V_+ = 0.41$V and $V_- = 0.55$V). A qualitatively similar asymmetry reduction was observed for the MTJs with 2.3nm of MgO barrier. Thus, we can conclude that in the junction with clean interfaces the two interfaces are of very different quality which leads to the strong asymmetric bias dependence of the TMR, consecutively leading to the inversion of the TMR at large positive bias when injecting towards the bottom atomically flat Fe/MgO interface. At the same time the presence of an O monolayer at this atomically flat Fe/MgO interface reduces the asymmetry of the TMR. Since that the addition of the O monolayer does not qualitatively change the bias dependence of TMR its role can be understood simply as an additional barrier. The fact that the asymmetry is more or less the same for all thicknesses suggests that it is predominantly determined by the interfaces.

### III. Model results

In order to gain more insight on how interface modifications affect the bias dependence of TMR we perform model calculations. We consider the standard two-probe setup, consisting of the scattering region ($S$) coupled to the left ($L$) and the right ($R$) ferromagnetic leads. Except that in our case we further subdivide the scattering region in three different parts $S = B_L|B|B_R$ corresponding to the parts adjacent to the left interface, middle, and adjacent to the right interface respectively. The electronic structure is described on the level of a single-orbital tight binding (TB) model. In the case of Fe/MgO-based MTJs, this model has predictive power due to the particular band structure of the ferromagnetic leads which can be thought to contain only one exchange-split $\Delta_1$ band.[29-35] This model represents an approximation in couple of ways: (i) it ignores the contribution to the transport of the bands with symmetries other than $\Delta_1$ and (ii) does not account for the contribution of the interface resonance state present in the minority Fe. It has been demonstrated previously that it describes remarkably well the behavior, not only the charge current and TMR behavior[29-31] but also the spin current and the spin-transfer torque (STT).[32-35] At the same time the model gives us the opportunity to readily calculate the voltage dependence of the current and TMR in MTJs with a variety of barrier modifications.

The Hamiltonian for the MTJ is

$$H = H_L + H_R + H_S + (H_{SL} + H_{SR} + h.c.)$$

where the different terms represent the isolated left and right leads $H_{L(R)} = \sum \varepsilon_\lambda^\sigma c_\lambda^{\sigma\dagger} c_\lambda^\sigma + \sum t_{\lambda\mu} c_\lambda^{\sigma\dagger} c_\mu^\sigma$, the scattering region $H_S = \sum \varepsilon_i c_i^{\sigma\dagger} c_i^\sigma + \sum t_{ij} c_i^{\sigma\dagger} c_j^\sigma$, and the coupling between the leads with the scattering region $H_{SL(R)} = \sum t_{a\alpha(b\beta)} c_{a(b)}^{\sigma\dagger} c_{\alpha(\beta)}^\sigma$. By convention the first principal layers of the left (right) interface are labeled $\alpha$ ($\beta$) indices in the leads and $a$ ($b$) in the barrier. The spin-split $\Delta_1$ bands in the Fe electrodes are represented by their spin-dependent on-site energies $\varepsilon^\uparrow = 3.0$ eV and $\varepsilon^\downarrow = 5.6$eV respectively. The on-site energy in the bulk MgO barrier is $\varepsilon_B = 9.0$eV ($B$ region). We vary the on-site energies in the $B_L$ and $B_R$ regions to model MTJs with modified interfaces. The nearest neighbor hopping matrix elements are chosen to be $t = -1.0$ eV in all regions. The Fermi energy is at $E_F = 0$eV. The calculations were carried out at low temperature.

The finite bias tunneling current is calculated using the standard Landauer formalism[36,37]

$$I_{\sigma\sigma'} = \frac{e}{h} \int dE d\mathbf{k}_\| [f(E, \mu_L) - f(E, \mu_R)] T_{\sigma\sigma'}(E, \mathbf{k}_\|)$$

where the transmission probability $T_{\sigma\sigma'}(E, \mathbf{k}_\|) = \text{Tr}[\Gamma_L^\sigma G^{\sigma\sigma'} \Gamma_R^{\sigma'} G^{\sigma\sigma'\dagger}]$ is integrated over $\mathbf{k}_\|$ in the surface Brillouin zone and over energy within the bias window. The bias window is controlled by the Fermi-Dirac distribution function $f$ where $\mu_{L(R)}$ is the chemical potential in the $L$ ($R$) lead. The retarded Green's function (GF) of the scattering region connected to the leads is $G^{\sigma\sigma'} = (g_S^{-1} - \Sigma_L^\sigma - \Sigma_R^{\sigma'})^{-1}$, where $g_S$ is the GF of the isolated scattering region and $\Sigma_{L(R)}^\sigma = t_{a\alpha(b\beta)}^2 g_{L(R)}^\sigma$ is the self-energy associated with the connection to the electrodes. The escape rate to the electrodes $\Gamma_{L(R)}^\sigma = i(\Sigma_{L(R)}^\sigma - \Sigma_{L(R)}^{\sigma\dagger}) = 2\pi t_{a\alpha(b\beta)}^2 \rho_{L(R)}^\sigma$ is proportional to the spin-dependent electrode surface density of states $\rho_{L(R)}^\sigma = -\text{Im}(g_{L(R)}^\sigma)/\pi$ (DOS). Within the single-band TB model and in the *limit of a thick barrier*, the transmission probability can be simplified to

$$T_{\sigma\sigma'}(E, \mathbf{k}_\|) = \frac{\Gamma_L^\sigma \Gamma_R^{\sigma'} |g_{ab}|^2}{|1 - g_{aa}\Sigma_L^\sigma|^2 |1 - g_{bb}\Sigma_R^{\sigma'}|^2}$$

where the assumption of a thick barrier allows for the multiple scattering terms in the denominator to be ignored. The interface transmission functions or probabilities (ITFs) can be expressed as $t_L^\sigma = (\Gamma_L^\sigma/t_{a\alpha}^2)(1 - t_{a\alpha}^2 g_{aa}^2)/|1 - g_{aa}\Sigma_L^\sigma|^2$ and $t_R^{\sigma'} = (\Gamma_R^{\sigma'}/t_{b\beta}^2)(1 - t_{b\beta}^2 g_{bb}^2)/|1 - g_{bb}\Sigma_R^{\sigma'}|^2$ by performing the wave function matching at each interface.[38] The ITFs have the meaning of the induced electrode DOS in the scattering region and can be written as a product of electrode's surface DOS and a spin-dependent function of the barrier potential at the interface $t_{L(R)}^\sigma = \rho_{L(R)}^\sigma D_{L(R)}^\sigma$. Clearly in the limit of a very high barrier $g_{ii} \ll 1$ and the ITFs reduce to the electrode DOS, which is the Jullière limit. Using the ITFs we can express the transmission probability as

$$T_{\sigma\sigma'}(E, \mathbf{k}_\|) = t_L^\sigma |S_{ab}|^2 t_R^{\sigma'}$$

where $S_{ab}$ can be interpreted as the matrix element of the scattering matrix across the barrier. This expression reduces to $T_{\sigma\sigma'} = t_L^\sigma e^{-2\kappa d} t_R^{\sigma'}$, where $\kappa$ is the decay constant in the barrier, at zero bias and a uniform barrier.[39-41] Finally TMR = $(I_P - I_{AP})/I_{AP}$ where $I_P = I_{\uparrow\uparrow} + I_{\downarrow\downarrow}$ and $I_{AP} = I_{\uparrow\downarrow} + I_{\downarrow\uparrow}$ are the currents for magnetizations in the electrodes parallel (P) and antiparallel (AP) to each other. It can be expressed through the ITFs as follows

$$\text{TMR} = \frac{\int d\varpi |S_{ab}|^2 (t_L^\uparrow t_R^\uparrow + t_L^\downarrow t_R^\downarrow - t_L^\uparrow t_R^\downarrow - t_L^\downarrow t_R^\uparrow)}{\int d\varpi |S_{ab}|^2 (t_L^\uparrow t_R^\downarrow + t_L^\downarrow t_R^\uparrow)}$$

where $\int d\varpi = \int dE \int d\mathbf{k}_\parallel [f(E, \mu_L) - f(E, \mu_R)]$. At zero bias and at the $\Gamma$ point only, it reduces to the familiar Jullière formula TMR = $2P_L P_R/(1 - P_L P_R)$, where $P_{L(R)} = (t_{L(R)}^\uparrow - t_{L(R)}^\downarrow)/(t_{L(R)}^\uparrow + t_{L(R)}^\downarrow)$ is the spin polarization of the DOS induced in the barrier by the electrodes.

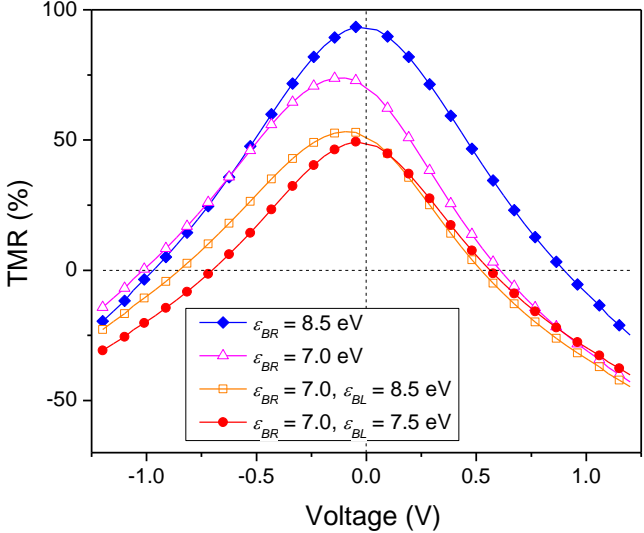

**Figure 2**: Calculated bias dependence of TMR in Fe//MgO/Fe and Fe/O/MgO/Fe MTJs. Oxygen in Fe/O/MgO/Fe MTJs is represented by 1 monolayer at the left interface with on-site energy $\boldsymbol{\varepsilon_{B_L}}$. The rough right interface is represented with 1 monolayer with onsite energy $\boldsymbol{\varepsilon_{B_R}}$. The thickness of the middle part of the MgO barrier is 4 monolayers.

Using this model we calculate the tunneling current and TMR in Fe/MgO/Fe MTJs corresponding to the experimental setup. For sample $A$ we assume that at the sharp interface the barrier is identical to the bulk $\varepsilon_{B_L} = \varepsilon_B = 9.0$ eV, while at the rough interface the barrier height is reduced. We vary $6.0 < \varepsilon_{B_R} < 9.0$eV which still represents a barrier layer albeit smaller than the bulk. The calculated TMR as a function of the applied bias is shown in Fig. 2. It can be seen that when $\varepsilon_{B_R}$ is close to the bulk value, the TMR is symmetric and close to the ideal case. As $\varepsilon_{B_R}$ decreases, the maximum value of TMR decreases and the asymmetry is increased. Moreover, we see that for large enough bias TMR changes sign, but due to the asymmetry this inversion happens for much smaller bias in the one direction than in the other. These results show qualitative agreement with the experimental observations. An asymmetry coefficient approximately equal to the experimental is obtained for the value of $\varepsilon_{B_R} = 7.0$eV. Thus, this model allows us to interpret many similar experiments in which one of the interfaces is morphologically different.[16-19] In those cases, disorder and roughness make the barrier at the interface more diffuse, effectively lowering the barrier height, which gives rise to the asymmetric behavior.

Similarly, we calculate the bias dependence of the TMR in sample $B$, where we fix $\varepsilon_{B_R} = 7.0$eV to account for the rough interface and vary the onsite energy of the O monolayer $7.0 < \varepsilon_{B_L} < 9.0$eV. The results, plotted in Fig. 2, show that the asymmetry is reduced. The same asymmetry as in experiment ($\gamma = 28\%$ for 1.1nm of MgO barrier) is obtained for $\varepsilon_{B_L} = 7.5$eV. Thus, within our model, the O layer can be well understood as an additional barrier. This is consistent with the experimental observation that the O layer attenuates both the P and AP spin-channels the same, which behavior is consistent with an additional barrier at the interface.[14]

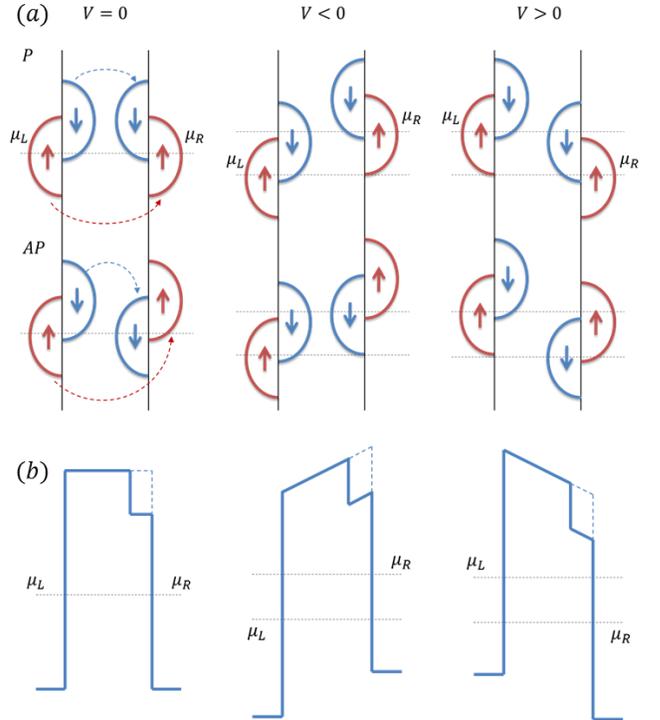

**Figure 3**: Schematic representation of the electronic structure of the electrodes and the barrier in a MTJ at different voltages. (a) Spin dependent DOS in Fe electrodes for zero and finite bias voltage. Vertical lines represent the left and right interface. The two spin channels correspond to transmission between inner-inner and outer-outer states. (b) Potential profile of asymmetric MTJs under negative and positive bias voltage. Negative (positive) bias decreases (increases) the barrier transparency.

In the case when all parts of the barrier are insulating the ITFs are largely determined by the DOS at the electrodes. Thus, the decrease in TMR and the subsequent sign reversal at finite bias could be qualitatively explained in terms of the realignment majority- and minority-spin DOS in the left and right electrodes.[8] The band alignments at zero and finite basis are illustrated in Fig. 3a. At zero bias the majority and minority bands are aligned in the P configuration, and misaligned in the AP configuration (due to exchange splitting). Thus, the majority channel in the P configuration ($I_{\uparrow\uparrow}$) dominates the current and TMR is positive ($I_P > I_{AP}$). Applied bias rigidly shifts the bands in the electrodes. The effect is that in the P configuration the bands get less and less aligned reducing $I_P$, while the alignment in the AP configuration improves. When the bias is comparable to the exchange splitting it aligns the majority and minority bands and the transmission probability becomes dominated one of the channels in the AP configuration ($I_{\downarrow\uparrow}$ for negative bias or $I_{\uparrow\downarrow}$ for positive bias) and TMR becomes negative ($I_P < I_{AP}$).

The behavior of TMR, in turn, can be related to the asymmetry of the $t_L$ and $t_R$ ITFs. The potential profiles of a barrier with an insulating layer of lower barrier height at the left interface are drawn in Fig. 3b for zero and finite bias. Applying negative bias decreases the effective barrier height at the left interface and increases it at the right interface by the same amount. Since on the right side the barrier is lower this reduction proportionately affects $t_R$ more than it affects $t_L$. Therefore, the overall transparency of the barrier is decreased. Positive bias does the opposite and increases the transparency of the barrier. As the result, compared to the symmetric barrier, smaller positive voltage is required to change the sign of TMR. However, higher negative voltage is necessary to do the same.

interface (green curve), one/two monolayers away from the left interface towards the center of the barrier (red/blue curve).

These observations allow us to generalize the model to explain experiments in which extra insulating layers are inserted at the interface, in particular NiFe/Ta$_2$O$_5$/Al$_2$O$_3$/NiFe[8] and CoFe/NiO/MgO/CoFe MTJs.[9] In both cases, asymmetric bias dependence of TMR and sign TMR inversion at high bias were observed. For example, to model the NiO/MgO composite barrier, we choose $\varepsilon_{B_R} = 9.0$eV the same as the bulk and $\varepsilon_{B_L} = 7.0$eV because the band gap of NiO is smaller compared to the MgO. The calculated bias dependence of TMR is shown in Fig. 4. TMR is symmetric for unmodified MgO barriers. The insertion of a thin NiO layer at the left interface introduces asymmetry in the bias dependence of TMR. Thus, for positive (negative) bias higher (lower) voltage is required to change the sign of TMR. However, placing the NiO layer inside the MgO barrier essentially restores the symmetry in bias dependence of TMR. When the NiO layer is at the interface $t_L$ is different from $t_R$ resulting asymmetric bias dependence of TMR. However a slight shift of the NiO layer from the interface yields $t_L = t_R$ and almost negligible asymmetry of TMR comes only through the asymmetric weighting with the S-matrix in the integral. Thus, the observed bias dependence and inversion of TMR can be understood simply in terms of asymmetric ITFs and the exchange bias in the electrodes.

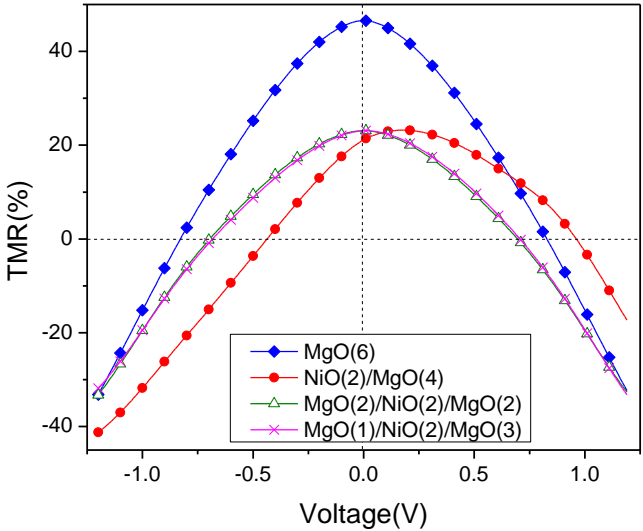

**Figure 4**: Bias dependence of TMR in CoFe/MgO/CoFe and CoFe/NiO/MgO/CoFe MTJs. The total barrier thickness is 6 monolayers for all curves. Black curve is for CoFe/MgO/CoFe MTJs. Green, blue and red curves are for CoFe/NiO/MgO/CoFe MTJs where the thickness of NiO is 2 monolayers. NiO is at the left

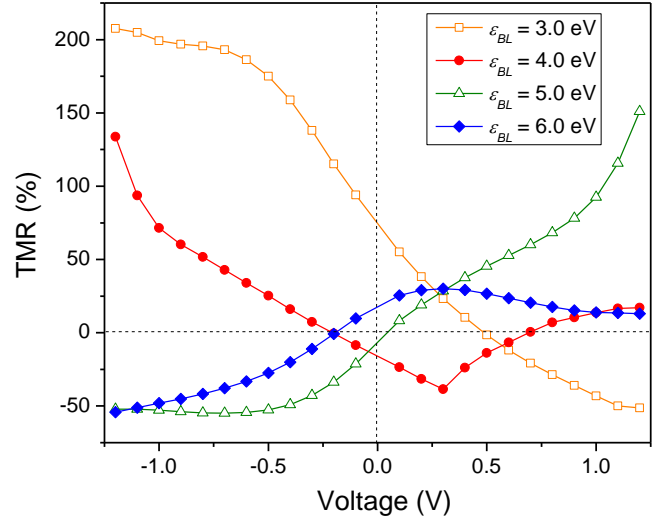

**Figure 5**: Bias dependence of TMR in MTJs in the resonant regime with a metallic layer of 1 monolayer thickness at the left interface. The total barrier thickness is 8 monolayers.

Further on, the model can be used to explain the bias dependence of TMR in MTJs with metallic interface layers, such as Fe/Cr/MgO/Fe[10] and Fe/Fe$_3$O$_4$/MgO/CoMTJs.[11] We model metallic layers by lowering the onsite energy of the interface layer $\varepsilon_{B_L} < 6.0$eV in which case the band crosses the Fermi level. The results for TMR are shown in Fig. 5 for several voltages. In the tunneling regime TMR has the 'usual' form with a maximum at zero voltage behaves as an even function of the voltage. For a shallow potential well at the interface ($\varepsilon_{B_L} = 5.0$ eV) the bias dependence of TMR already shows qualitatively different behavior, TMR inversion cannot be achieved by applying positive voltage. In this case we notice the interesting feature that TMR becomes and odd function of the bias. This suggests a possibility to combine memory and logic functions in the same bit.[42,43] Moreover, TMR increases with voltage. In general memory stacks are operated at low voltage because TMR has a maximum. This feature could allow memories to be operated at higher voltage. Similar behavior but with a reverse sign is observed for $\varepsilon_{B_L} = 3.0$ eV. Finally, for $\varepsilon_{B_L} = 4.0$ eV TMR completely reverses its behavior. It is inverse at zero bias and increases with applied bias. Similar qualitative behavior was observed experimentally in Fe/Fe$_3$O$_4$/MgO/Co MTJs.[11]

These features can be explained by the appearance of resonance states in the potential well created by the metallic layer at the interface which contribute to resonant tunneling.[44] In this regime it becomes possible that $|1 - g_{aa}\Sigma_L^\sigma| \approx 0$ for some combinations of $E$ and $\boldsymbol{k}_\parallel$. The position of the resonance also depends parametrically on $\varepsilon_{B_L}$. At resonance, the ITF at the left interface $t_L^\sigma = \rho_L^\sigma D_L^\sigma$ will be strongly enhanced. The behavior of the TMR with voltage will depend strongly on the way resonances overlap with the DOS of the electrodes.

To illustrate this we plot, in Fig. 6, the zero bias $\boldsymbol{k}_\parallel$-resolved DOS for the electrode (representative of $t_R$) and the first barrier layer for $\varepsilon_{B_L} = 3.0$ and $\varepsilon_{B_L} = 4.0$ eV (representative of $t_L$). Both cases there is one resonance in each spin channel. For $\varepsilon_{B_L} = 3.0$ eV the resonance overlaps with the majority DOS but it does not with the minority DOS. Thus, the term $t_L^\uparrow t_R^\uparrow$ gives the largest contribution into current and there is only non-resonant transmission in the AP configuration. Thus, $I_P > I_{AP}$ and TMR is positive. For $\varepsilon_{B_L} = 4.0$ eV the minority resonance starts to overlap with the electrode DOS, which results in very strong AP transmission, $I_{AP} > I_P$, and TMR becomes negative. As $\varepsilon_{B_L}$ increases the resonances disappear altogether and we have a transition to the insulating regime at $\varepsilon_{B_L} = 6$eV. Conversely for lower $\varepsilon_{B_L}$ or for thicker interface more than one resonance can appear.

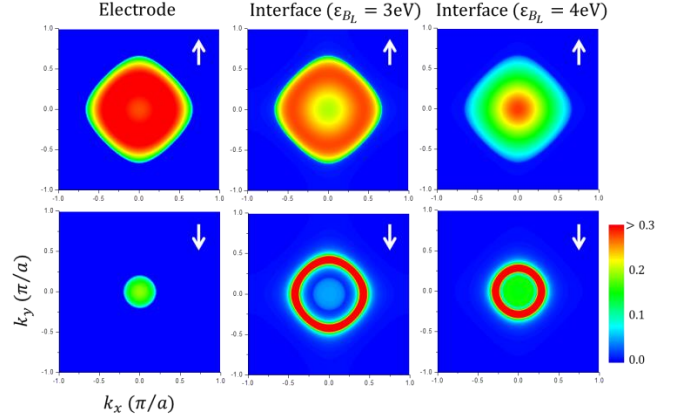

**Figure 6**: $\boldsymbol{k}_\parallel$-resolved majority and minority DOS in the electrodes (left panels), and induced DOS in the first barrier layer at the left interface for $\varepsilon_{B_L} = 3$eV (middle panels) and $\varepsilon_{B_L} = 4$eV (right panels).

### V. Conclusions

We report asymmetric bias dependence of TMR and TMR reversal at large bias in single crystal Fe/MgO based MTJs: one with two morphologically different interfaces and one with a layer of O inserted at the bottom interface. We develop a model to describe the TMR behavior in MTJs with modified barriers, with the added benefit that is allows us to reinterpret and categorize the wide variety of experimental results. The model, although simple, captures the essential physics of the Fe/MgO-based MTJs (namely the $\Delta_1$ filtering) and agrees qualitatively with the experimental data. It predicts two distinct regimes: tunneling when the modified interfaces are insulating and resonant when the interfaces become metallic. In the tunneling regime, we observe asymmetric bias dependence of the TMR resulting from the asymmetry of the ITFs due to difference of the morphology of the interfaces. We explain the TMR inversion results from the realignment of the electrode DOS in large bias, which overcomes the exchange splitting in the electrodes. In the resonant regime, the electron tunneling through the resonant levels at the interface can dominate the transmission probability for a particular spin channel. Judicious choices of the position of the resonant level allow for control the shape of the TMR bias dependence. We have demonstrated cases of normal TMR decreasing with bias, inverse TMR increasing with bias, and TMR changing sign for positive and negative bias.

### Acknowledgements

The work at the University of Puerto Rico was supported by NSF (Grants Nos. EPS-1002410, EPS-1010094, and DMR-1105474) and DOE (Grant No.DE-FG02-08ER46526). P.-J. Zermatten, G. Gaudin, F. Bonell, S. Andrieu, and C.Tiusan acknowledge the SPINCHAT project ANR-07-BLAN-341. C. Tiusant acknowledges POS CCE ID. 574, code SMIS-CSNR 12467.